\documentclass[10pt,aps,prl,twocolumn,showpacs,showkeys,superscriptaddress,longbibliography]{revtex4-2}

\usepackage{mathtools,bm,amssymb,accents,hyperref,textcomp,siunitx}

\newcommand*\sfrac[2]{{}^{#1}\!/_{#2}}

\newlength{\dhatheight}
\newcommand{\hhatTS}[1]{
    \settoheight{\dhatheight}{\ensuremath{\hat{#1}}}
    \addtolength{\dhatheight}{-0.2ex}
    \hat{\vphantom{\rule{1pt}{\dhatheight}}
    \smash{\hat{#1}}}}    
\newcommand{\hhatS}[1]{
    \settoheight{\dhatheight}{\ensuremath{\scriptstyle{\hat{#1}}}}
    \addtolength{\dhatheight}{-0.15ex}
    \hat{\vphantom{\rule{1pt}{\dhatheight}}
    \smash{\hat{#1}}}}
\newcommand{\hhatSS}[1]{
    \settoheight{\dhatheight}{\ensuremath{\scriptscriptstyle{\hat{#1}}}}
    \addtolength{\dhatheight}{-0.075ex}
    \hat{\vphantom{\rule{1pt}{\dhatheight}}
    \smash{\hat{#1}}}}
\newcommand{\hhat}[1]{\mathchoice{\hhatTS{#1}}{\hhatTS{#1}}{\hhatS{#1}}{\hhatSS{#1}}}

\begin{document}

\title{Second order phase dispersion by optimised rotation pulses}
\author{David L. Goodwin}
\email{david.goodwin@chem.ox.ac.uk}
\affiliation{Institute for Biological Interfaces 4 -- Magnetic Resonance, Karlsruhe Institute for Technology (KIT), Karlsruhe, Germany}
\affiliation{Chemistry Research Laboratory, University of Oxford, Mansfield Road, Oxford, OX1 3TA, UK}
\author{Martin R. M. Koos}
\affiliation{Institute for Biological Interfaces 4 -- Magnetic Resonance, Karlsruhe Institute for Technology (KIT), Karlsruhe, Germany}
\affiliation{Department of Chemistry, Carnegie Mellon University, Pittsburgh, USA}
\author{Burkhard Luy}
\affiliation{Institute for Biological Interfaces 4 -- Magnetic Resonance, Karlsruhe Institute for Technology (KIT), Karlsruhe, Germany}
\affiliation{Institute of Organic Chemistry, Karlsruhe Institute for Technology (KIT), Karlsruhe, Germany}

\date{\today}

\begin{abstract}
We show that the duration of broadband universal control pulses can be halved by choosing control targets with a quadratic function of phase dispersion. This class of control pulses perform a broadband universal rotation around an axis, in the Bloch sphere representation of two-level systems, given by this phase dispersion function. We present an effective optimal control method to avoid the problem of convergence to local extrema traps.
\end{abstract}




\maketitle

\section{Introduction}

The critically important milestones on the road-map of quantum technology development are expected to include quantum optimal control \cite{Glaser:2015,Acin:2018}. The main result of this communication is to reduce the time and energy required to control a quantum system. As an application of optimal control theory, quantum optimal control can be tasked with finding control pulses, usually in the form of electromagnetic radiation, to perform a desired operation on an arbitrary quantum state. Already, this method has been applied successfully to a wide range of experiments including magnetic resonance spectroscopy \cite{Khaneja:2005,Tosner:2006}, error-correction for quantum computing \cite{Zhang:2012}, quantum information registers \cite{Dolde:2014}, robust atom interferometry \cite{Saywell:2018,Saywell:2020}, and high-resolution medical imaging satisfying legal irradiation constraints \cite{Vinding:2017}.

The growing number of reported applications of quantum optimal control are expanded by the concurrent development of optimal control algorithms. There are a number of approaches to optimal control, such as Lagrangian methods \cite{Somloi:1993}, time optimal control \cite{Khaneja:2001}, annealing quantum systems to a desired effective Hamiltonian \cite{Tosner:2006}, sophisticated gradient-free searches \cite{Doria:2011}, rapidly converging Krylov-Newton methods \cite{Ciaramella:2015}, and optimal control using analytic controls \cite{Machnes:2018}. This development is essential as desired solutions push to the limit of what is physically possible and control problems become computationally, and numerically challenging \cite{Pechen:2011}.

Quantum optimal control problems can be solved numerically by using a piecewise-constant control pulse approximation \cite{Conolly:1986,Mao:1986,Viola:1998,Viola:1999}, and the trajectory gradient can be calculated and followed using the \emph{gradient ascent pulse engineering} (\textsc{grape}) method \cite{Khaneja:2005}. The initial \textsc{grape} method showed linear convergence to a desired optimum destination, and was later extended to give superlinear convergence \cite{deFouquieres:2011}, then quadratic convergence \cite{deFouquieres:2012,Goodwin:2016}. This communication introduces an ordered optimal control method, based on the \textsc{grape} method, to avoid local convergence traps encountered in the optimal control problem set out below. This method is termed \emph{morphic}-\textsc{grape} by the authors.

In addition, and vital to effective experimental applications, optimal control has advanced theoretical studies calculating universal gates for quantum computing \cite{Viola:1999b,SchulteHerbruggen:2005,Grace:2007,Goerz:2014,Zahedinejad:2015}. A subset of universal quantum gates include control targets designed as an effective rotation in the Bloch sphere representation of a two-level system, with a defined rotation axis and rotation angle, for any initial basis-state vector \cite{Kobzar:2012,Dolde:2014}. The developments reported here will be concerned with this \emph{universal rotation} class of control pulses.

One of the drawbacks associated with universal rotation solutions is the long duration of control pulses when compared with the easier control problem of optimising \emph{state-to-state} problems \cite{Kobzar:2012,Glaser:2015}. The optimal control method presented here will show that the previously required pulse duration can be halved by defining target propagators as a function of phase dispersion.

\section{Broadband optimal control to effective propagators}

An ensemble of noninteracting two-level systems should include frequency dispersion terms from local environmental conditions. Uniform manipulation of the ensemble is a difficult practical task but by describing the ensemble as a bilinear control problem \cite{Conolly:1986,Viola:1998}, numerical optimisation can be used to find solutions that are difficult to find analytically. In formulating this bilinear control problem, the controllable parts of the Hamiltonian are the pulses, and the uncontrollable part is the local frequency dispersion term. Restricting control pulses to phase modulation, $\varphi(t)$, with constant amplitude $A$, the time-dependent Hamiltonian for this ensemble of $K$ noninteracting two-level systems can be written as
\begin{equation}
 \hat{H}(t)=\sum\limits_{k=1}^{K}\omega_k^{}\hat{\sigma}_{\text{z}}^{(k)}+A\cos{(\varphi(t))}\hat{\sigma}_{\text{x}}^{(k)}+A\sin{(\varphi(t))}\hat{\sigma}_{\text{y}}^{(k)}
 \label{EQ_BilinearHamiltonian}
\end{equation}
where the angular frequency $\omega_k^{}$ is variously termed the resonant frequency offset, chemical shift, or detuning, and describes the local frequency dispersion within the ensemble. Practically, this ensemble of resonant frequency offsets form a discrete grid of noninteracting two-level systems spread over a relevant bandwidth \cite{Skinner:2003,Daems:2013}. $\hat{\sigma}_\text{x,y,z}^{(k)}$ are operators of the $k^\text{th}$ two-level system, operating in a Hilbert space and related to the Pauli matrices $\hat{\sigma}_\text{x,y,z}$ by
\begin{equation}
\hat{\sigma}_\text{x,y,z}^{(k)}=E_{}^{(k)}\otimes\hat{\sigma}_\text{x,y,z}
\end{equation}
where $E_{}^{(k)}$ is a $K\times K$ single-entry matrix with the diagonal element $E_{kk}^{(k)}=1$.

The \textsc{grape} (gradient ascent pulse engineering) method of optimal control \cite{Khaneja:2005} proceeds to describe the time-dependent control pulses as piecewise constant over a small time interval $\Delta t$ \cite{Viola:1998,Viola:1999}. This approximation allows for a numerical solution of Eq.~(\ref{EQ_BilinearHamiltonian}) through time-ordered propagation,
\begin{align}
 & \hhat{P}_{\!n}=\exp{\big[-i\hhat{L}_n\Delta t\big]}, & \begin{array}{l}\mathcal{U}_n=\hhat{P}_{\!n}\dots\hhat{P}_{\!2}\hhat{P}_{\!1} \\ \mathcal{V}_n=\hhat{P}_{\!n+1}^{\dagger}\dots\hhat{P}_{\!N}^{\dagger}\mathcal{R}\end{array} &
 \label{EQ_EffectivePropagators}
\end{align}
with the Liouvillian $\hhat{L}_n$ being an adjoint representation of a Hamiltonian: $\hhat{L}=\openone\otimes\hat{H} - \hat{H}^\dagger\otimes\openone$ \cite{deFouquieres:2011}. At a time increment $n$, the effective propagator $\mathcal{U}_n$ evolves the system forward in time over the interval $[0,t_n]$, and the effective propagator of the adjoint control problem, $\mathcal{V}_n$, evolves backwards from a desired target $\mathcal{R}$ over the interval $[T,t_n]$. The effective propagator $\mathcal{U}_N$ is the solution to the ensemble Hamiltonian Eq.~(\ref{EQ_BilinearHamiltonian}) over the total pulse duration $T$. The effective propagator of the adjoint problem, $\mathcal{V}_n$, is needed for a \emph{fidelity gradient} calculation, outlined below.

The Hilbert-Schmidt inner product of the desired effective propagator $\mathcal{R}$ and the effective propagator over the control duration, $\mathcal{U}_N$, gives a measure to numerically maximise and is termed the fidelity \cite{Khaneja:2005,Tosner:2006,Kobzar:2012,Dolde:2014}:
\begin{equation}
 \max_{\varphi}\big\{\mathcal{F}\big\}=\frac{1}{d}\max_{\varphi}\Big\{\text{Re}\big\langle\mathcal{R}\big|\mathcal{U}_N^{}\big\rangle\Big\}
 \label{EQ_maxFidelity}
\end{equation}
This fidelity measure $\mathcal{F}$ is normalised by the dimension of the Hilbert space $d$ to give sensible and predictable bounds $\mathcal{F}\in[-1,+1]$. The desired target $\mathcal{R}$ is interpreted as a rotation in the Bloch sphere representation in all that follows.

Given an initial guess for control pulses, optimal control pulses can be found by stepping in the direction of the fidelity gradient vector over control manifold, until a maximum fidelity is found. The distance to step is set by an appropriate Newton-type optimisation method \cite{Goodwin:Thesis}. The gradient-following \textsc{grape} method \cite{Khaneja:2005} requires a fidelity gradient vector $\nabla \mathcal{F}$ and, for phase modulated control, this is constructed from the elements
\begin{equation}
 \frac{\partial \mathcal{F}}{\partial \varphi_{n}^{}}=\text{Re}\big\langle\mathcal{V}_n\big|\Bigg[A\cos{(\varphi_n^{})}\frac{\partial\hhat{P}_{\!n}}{\partial\hhat{\sigma}_\text{y}}-A\sin{(\varphi_n^{})}\frac{\partial\hhat{P}_{\!n}}{\partial\hhat{\sigma}_\text{x}}\Bigg]\mathcal{U}_n\big\rangle
 \label{EQ_DirectionalDerivatives}
\end{equation}
The two partial derivatives on the right-hand side are directional propagator derivatives, and can be interpreted at the derivative of the propagator, $\hhat{P}_{\!n}$, in the direction of the operators $\hhat{\sigma}_\text{x}$ and $\hhat{\sigma}_\text{y}$. The fidelity gradient is calculated by using the same fidelity definition of Eq.~(\ref{EQ_maxFidelity}), with the directional propagator derivatives sandwiched by the effective propagator and its adjoint in Eq.~(\ref{EQ_EffectivePropagators})

For each time slice, the time propagator in Eq.~(\ref{EQ_EffectivePropagators}) and each of the directional derivatives in Eq.~(\ref{EQ_DirectionalDerivatives}) can be calculated analytically with one exponentiation of a block-triangular matrix \cite{VanLoan:1978,Goodwin:2015},
\begin{equation}
\exp{\Bigg[-i\begin{bmatrix} \hhat{L}_n^{} & \hhat{\sigma} \\ \bm{0} & \hhat{L}_n^{} \end{bmatrix}\Delta t \Bigg]}=\begin{bmatrix} \hhat{P}_{\!n} & \dfrac{\partial\hhat{P}_{\!n}}{\partial\hhat{\sigma}} \\ \bm{0} & \hhat{P}_{\!n} \end{bmatrix}
\label{EQ_Auxmat}
\end{equation}
in each direction $\hhat{\sigma}\in\{\hhat{\sigma}_\text{x},\hhat{\sigma}_\text{y}\}$. Considering the large and sparse Hamiltonian formulation of Eq.~(\ref{EQ_BilinearHamiltonian}), the main computational cost is $2N$ exponential operations, which can be performed in parallel and can include a efficient propagator recycling scheme \cite{Goodwin:2016}.

Pulses are linearly scaleable with the desired bandwidth \cite{Glaser:2015}, with the transform from one pulse $\text{p}_1$, with duration $T_1$, amplitude $A_1$, and covering a bandwidth $\Omega_1$, to a second pulse $\text{p}_2$, with duration $T_2$, amplitude $A_2$, and covering a bandwidth $\Omega_2$, having the same effect if
\begin{align}
& \Omega_1 T_1=\Omega_2 T_2=b && \text{and} && \frac{\beta}{A_1 T_1}=\frac{\beta}{A_2 T_2}=s &
\end{align}
where the subscripts refer to the according property of pulse $\text{p}_1$ and $\text{p}_2$. The dimensionless \emph{bandwidth factor} $b$ and \emph{scaling factor} $s$ are defined here to characterise the pulse, rather than the system-specific quantities of bandwidth, pulse amplitude, and pulse duration. This is common in the discipline of magnetic resonance, and the \emph{rotation angle} $\beta$ is included in the scaling factor to normalise with this use in magnetic resonance. In this definition, bandwidths are measured in Hertz, pulse duration in seconds, pulse amplitudes in radians per second, and rotation angles in radians. In the investigation that follows the scaling factor is chosen to be $s=\sfrac{2\beta}{\pi b}$ which, in effect, gives the same pulse amplitude over all bandwidth factors for a defined system.

The choice of constant amplitude, phase modulated, control pulses is one of inference: both chirped pulses \cite{Baum:1985,Kupce:1995,Garwood:2001,Jeschke:2015,Spindler:2017} and universal rotation pulses \cite{Kobzar:2012} are close to constant amplitude, and it is expected that pulses produced in this work will have a similar form.

The piecewise-constant time slices used in calculations are a very good approximation in nuclear magnetic resonance (\textsc{nmr}) spectroscopy. In cases where this piecewise-constant approximation is not valid, such as applications to electron paramagnetic resonance (\textsc{epr}) spectroscopy, a feedback control method \cite{Egger:2014,Goodwin:2018} or a transfer-matrix method \cite{Spindler:2012,Doll:2013,Kaufmann:2013} can be used to calibrate pulses due to these hardware-specific cavity effects. Further to this, optimal control can be designed to include a robustness to pulse amplitude miscalibration \cite{Skinner:2003}.

\section{Rotation axes with quadratic phase dispersion}

Derived from Landau-Zener-St\"{u}ckelberg-Majorana theory \cite{Jeschke:2015}, the ensemble of Eq.~(\ref{EQ_BilinearHamiltonian}) can be controlled with adiabatic evolution \cite{Meister:2014,Leghtas:2014,Augier:2018} and is described as a linear frequency sweep over the two-level systems. The phase dispersion resulting from a linear frequency sweep is quadratic, and can be used to define the local targets of an optimal control method. Previous work on this theme has optimised state-to-state problems with linear phase dispersion \cite{Gershenzon:2008,Koos:2015,Koos:2017} and quadratic phase dispersion \cite{Skinner:2010,Meister:2014,Brif:2014,Altenhof:2019}.

Although state-to-state solutions are useful, they are only functionally defined for a specified state preparation: the desired control is only effective for a given initial state of the system. A \emph{universal} control method, independent of the initial state, attempts to find desired unitary propagators which represent a rotation of the Bloch sphere; rotating all components of the coordinate system about an axis through the origin \cite{Kobzar:2012,Dolde:2014}.

The effective unitary propagator representing a universal rotation can be formulated for each member of the ensemble, $k$,
\begin{equation}
\mathcal{R}_k^{}(\beta)=\exp{\big[-i \hhat{n}_k^{}\beta\big]}
\label{EQ_TargetPropagators}
\end{equation}
for a rotation angle $\beta$. The desired axes of rotation, $\hhat{n}_k^{}$, can be defined as
\begin{equation}
\hhat{n}_k^{}=\cos{(\alpha_k^{})}\hhat{\sigma}_\text{x}^{(k)}+\sin{(\alpha_k^{})}\hhat{\sigma}_\text{y}^{(k)}
\end{equation}
where the angle $\alpha_k^{}$ is a phase on the transverse plane of the Bloch sphere and is described by a phase dispersion function. This study uses a quadratic phase dispersion similar to that of the linear frequency swept chirped pulses \cite{Baum:1985,Kupce:1995,Garwood:2001,Jeschke:2015,Spindler:2017}:
\begin{equation}
\alpha_k^{}\triangleq\pi bQ\bigg(1-\frac{\omega^2_{k}}{\Omega^2_{}}\bigg)
\label{EQ_PhaseDispersion}
\end{equation}
This is the phase that the transverse components of states acquire during the pulsing time \cite{Gershenzon:2008}. The introduction of $Q$ is defined here as the quadratic coefficient of a phase dispersion function, which is closely related to previous work using a linear phase dispersion function \cite{Gershenzon:2008}. The case when $Q=0$ has been previously published as Broadband Universal Rotation By Optimised Pulses (\textsc{burbop}) \cite{Kobzar:2012}.

It should be emphasised that the desired pulses in the communication are an improvement towards universal rotation pulses with reduced control duration, using a phase dispersion function similar to chirped pulses, rather than a generalisation of chirped pulses to universal rotation pulses. Furthermore, the function in Eq.~(\ref{EQ_PhaseDispersion}) is a quadratic function of $Q$ but could be formulated, in principle, by any function of $Q$. This new class of pulses are named Second ORder phase Dispersion by Optimised Rotation (\textsc{sordor}) pulses by the authors.

\section{Morphic optimal control}

\begin{figure}
\centering{\includegraphics{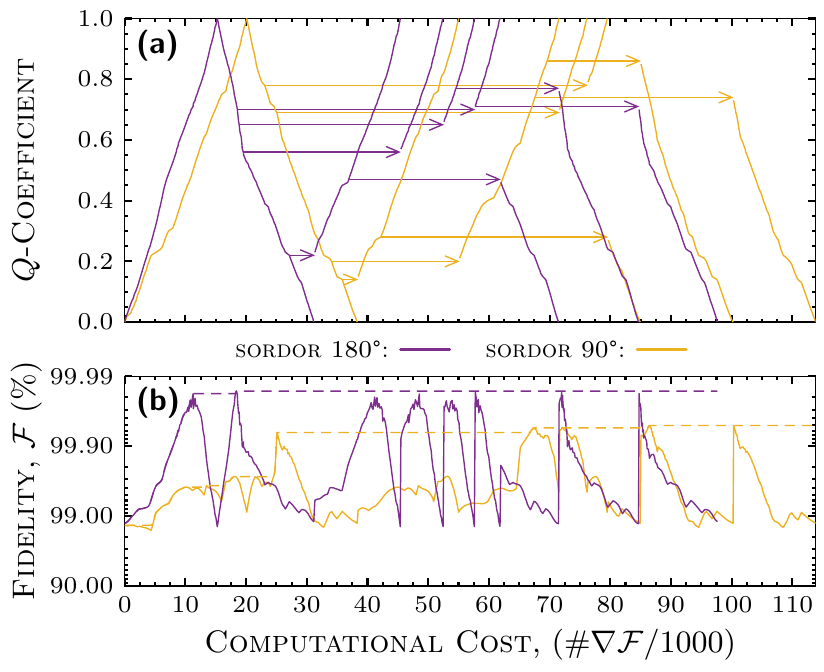}}
\caption{An indication of the computational cost of calculating \textsc{sordor} pulses at $b=18.0$. Panel (a) shows the number of fidelity gradient calculations as the $Q$-coefficient is increased and decreased, with the forward, backward and smoothing stages of this morphic optimal control method. Panel (b) shows the corresponding fidelity as a function of gradient calculations. Horizontal arrows in panel (a) show the point from which the waveform is morphed to start the smoothing stage. Dashed lines in panel (b) show the maximum fidelity obtained.}
\label{FIG_sordor_convergence}
\end{figure}

In the authors' experience, the standard procedure of starting \textsc{grape} from a random guess or an informed guess, \emph{e.g.} an adiabatic passage pulse, produces a fidelity profile periodically reducing to zero over the bandwidth, particularly when setting $Q\gtrapprox 0.1$ in Eq.~(\ref{EQ_PhaseDispersion}). The reason for this is that, as the optimisation progresses, fidelity sections either side of the zeroed fidelity increase, also increasing the average fidelity, trapping these local fidelity minima into the control manifold \cite{Pechen:2011,Goodwin:2016}. This can make the process of finding the desired optimal solutions set out in Eq.~(\ref{EQ_TargetPropagators}) a difficult task. Furthermore, a high average fidelity over the bandwidth is not sufficient for acceptable pulse performance -- the pulse should uniformly manipulate the entire bandwidth.

An ordered optimisation procedure of \emph{morphic}-\textsc{grape} is set out below to avoid these traps. This method is effective, yet easy to use, and proceeds by morphing one optimisation problem into an incrementally modified optimisation problem; using the solution of one as the new starting point of the next.

A grid of coefficients, $Q\in[0,1]$, and bandwidth factors, $b\in(0,18]$, sets a \emph{morph area} for the results that follow. \textsc{Sordor} pulses are created with four directional morphs: forward morph, incrementing $Q$ with $+\Delta Q$; backward morph, decrementing $Q$ with $-\Delta Q$; compressed morph, interpolating a control pulse from $b$ to $b-\Delta b$; and expanded morph, interpolating a control pulse from $b$ to $b+\Delta b$. An additional smoothing stage can be used to remove discontinuities in the fidelity profile, using sequences of forward or backward morphs from the largest extrema in the $\frac{\mathrm{d}\mathcal{F}}{\mathrm{d}Q}$ profile. The ordered recipe of finding \textsc{sordor} pulses is outlined as follows:
\begin{enumerate}
\item Start with a universal rotation pulse ($Q=0$) and use \emph{morphic}-\textsc{grape}:
\begin{enumerate}
\item Forward, increasing from $Q=0$.
\item Backward, decreasing from $Q=1$.
\end{enumerate}
\item Smoothing \emph{morphic}-\textsc{grape} for $b=18$, from extrema in the $\frac{\mathrm{d}\mathcal{F}}{\mathrm{d}Q}$ profile.
\item Start with a \textsc{sordor} pulse of high bandwidth factor ($b=18$) and use \emph{morphic}-\textsc{grape}:
\begin{enumerate}
\item Compressed, decreasing from $b=18$.
\item Expanded, increasing from $b=\Delta b$.
\end{enumerate}
\end{enumerate}
Empirical investigation finds an adequate grid spacing of $\Delta b=0.2$ and $\Delta Q=0.01$. Convergence is accepted when $\|\nabla \mathcal{F}\|\leqslant \min{\left(10^{-4},10^{-5\left(\frac{210}{583}+\frac{b}{36}\right)}\right)}$. The smoothing morph starts from the eleven largest extrema, with the first five of these shown in Fig.~\ref{FIG_sordor_convergence}. Figure \ref{FIG_sordor_convergence}(a) shows the places on the $Q$-coefficient grid where the smoothing starts, after the forward and backward stages, and Fig.~\ref{FIG_sordor_convergence}(b) shows how the fidelity increases and decreases over these stages. Both are plotted as a function of the number of gradient calculations which shows the computational cost of \emph{morphic}-\textsc{grape} at $b=18.0$, which can be regarded as the seed for the compressed and expanded stages. The forward and backward stages at different values of $b$ show a similar profile to those stages in Fig.~\ref{FIG_sordor_convergence}.

A direct comparison of the computational cost of this \emph{morphic}-\textsc{grape} method to the previously published universal rotation pulses \cite{Kobzar:2012} is not appropriate, because \emph{morphic}-\textsc{grape} essentially includes that previous work.

\section{Results}

\begin{figure}
\centering{\includegraphics{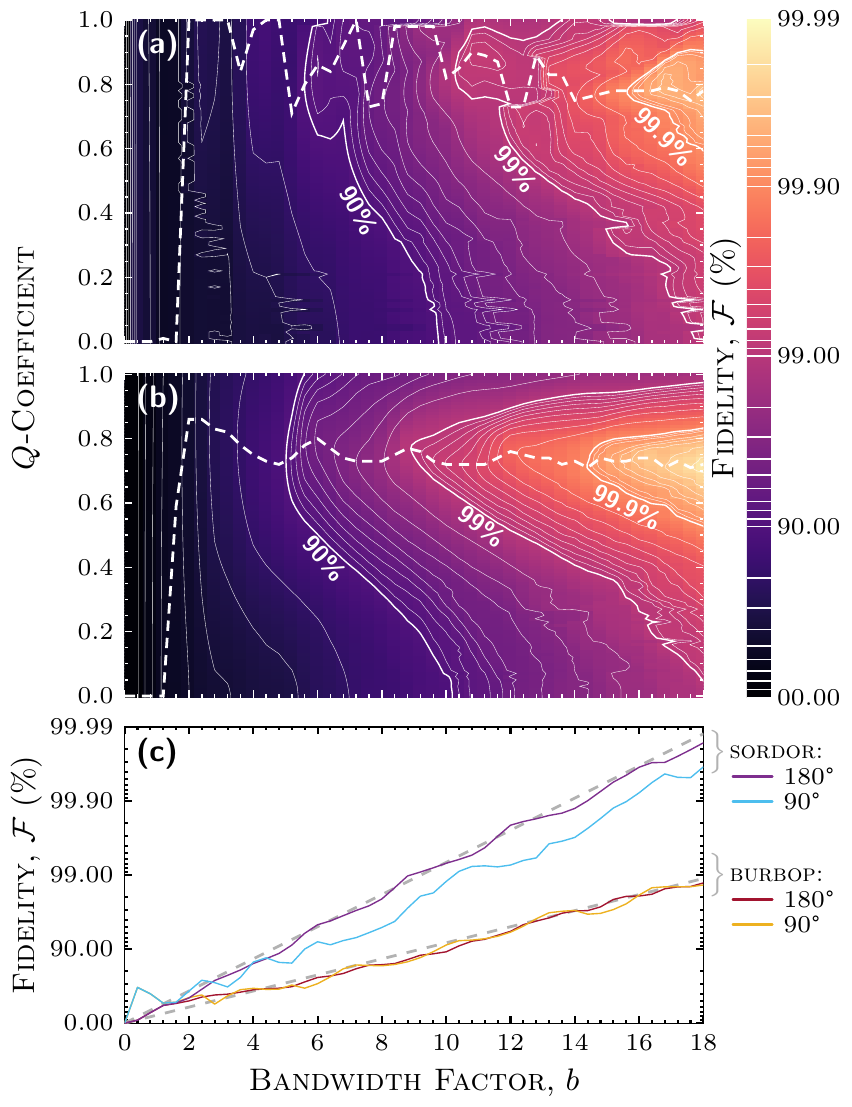}}
\caption{Fidelity of (a) \textsc{sordor} $\beta=\sfrac{\pi}{2}$ pulses, and (b) \textsc{sordor} $\beta={\pi}$ pulses, over a grid of $Q$-coefficients and bandwidth factors. The dashed lines in (a) and (b) are the maximum fidelities at each bandwidth factor. Plot (c) shows these maximum \textsc{sordor} fidelities compared with the previous best universal rotation pulses (\textsc{burbop}). Dashed lines in panel (c) are reference plots $\exp(-0.50b)$ and $\exp(-0.25b)$.}
\label{FIG_infidelity_duration_multiplier_1}
\end{figure}

The main results of this work are shown in Fig.~\ref{FIG_infidelity_duration_multiplier_1}. The \emph{Spinach} optimal control toolbox \cite{deFouquieres:2011,Goodwin:2015,Goodwin:2016} is used to simulate the ensemble of two-level systems, defined by Eq.~(\ref{EQ_BilinearHamiltonian}), with minimal modification to enable optimisation to the desired unitary propagators in Eq.~(\ref{EQ_TargetPropagators}). A linearly spaced resonance offset profile in Eq.~(\ref{EQ_BilinearHamiltonian}), with $K=1+\lceil10b\rceil$, and a piecewise constant approximation of Eq.~(\ref{EQ_EffectivePropagators}), with $N=50b$, is optimised with the $\ell$-\textsc{bfgs} method \cite{deFouquieres:2011}, forming a Hessian approximation from the previous twenty gradients.

It is not only interesting to see how the fidelity increases over the grid of $Q$-coefficients but it is also practically useful because experiments requiring combinations of pulses with different rotation angles $\beta$ should be \emph{matched}, \emph{i.e.} a sequence consisting of pulse $\text{p}_1$ followed by pulse $\text{p}_2$, should have $b_{\text{p}_1}=mb_{\text{p}_2}$ with $Q_{\text{p}_2}=mQ_{\text{p}_1}$, where $m$ is an arbitrary multiplier. The maximum fidelities found over the grid of $Q$-coefficients and bandwidth factors, from the forward, backward, compressed, and expanded \emph{morphic} optimisations, are shown in Figs.~\ref{FIG_infidelity_duration_multiplier_1}(a) and \ref{FIG_infidelity_duration_multiplier_1}(b).

It should be noted that the \emph{morphic} optimisations for $\beta=\sfrac{\pi}{2}$ are difficult, numerically and computationally; the backward, compressed, and expanded stages of the \emph{morphic} optimisations give little improvement when $\beta=\pi$, and were designed primarily for the case when $\beta=\sfrac{\pi}{2}$.

Figure \ref{FIG_infidelity_duration_multiplier_1}(c) shows the maximum fidelity at each bandwidth factor compared with the fidelity of an equivalent universal rotation pulse, \emph{broadband universal rotations by optimised pulses} (\textsc{burbop}): the shortest duration universal rotation pulses known to the authors \cite{Kobzar:2012,Glaser:2015}. The \textsc{burbop} in this comparison have the same constant amplitude $A$ and scaling factor $s$ as the \textsc{sordor} pulses. \textsc{Burbop} pulses start from the same ones as published previously \cite{Kobzar:2012}, then are further optimised with the same exact gradients [Eq.~(\ref{EQ_Auxmat})] and optimisation tolerances. It is clear that \textsc{sordor} pulses show significant improvement over \textsc{burbop}.

\section{Example of an application in magnetic resonance spectroscopy}

\begin{figure*}
\centering{\includegraphics{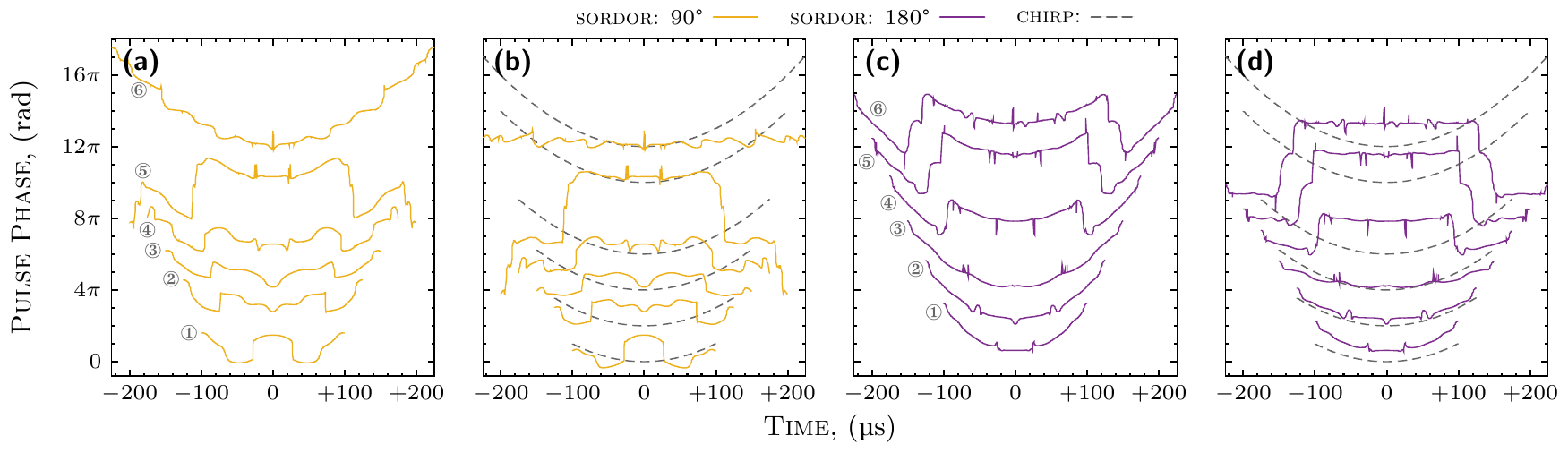}}
\caption{A selection of \textsc{sordor} pulses showing the unwrapped phase of the constant amplitude pulse as a function of time, measured from the centre of the pulse. (a) $\beta=\sfrac{\pi}{2}$ pulses at six pulse durations; (b) the same pulses shown with a quadratic phase subtracted; (c) $\beta={\pi}$ pulses at six different pulse durations; (d) the same pulses shown with a quadratic phase subtracted. The formula for the quadratic phase, shown as a dashed line in panels (b) and (c), is outlined in the main text.}
\label{FIG_sordor_pulses}
\end{figure*}

This section will proceed to apply the phase modulated pulses above, the subject of this article, to the experimentally relevant system of nuclear magnetic resonance spectroscopy.

As an example, a spin-$\sfrac{1}{2}$ $^{13}$C nucleus in a static magnetic field of $B_0=14.0954~\si{\tesla}$ forms the ensemble of two-level systems in Eq.~(\ref{EQ_BilinearHamiltonian}). For the analysis of small molecules, pulses should cover a bandwidth of $\Omega=37.5~\si{\kilo\hertz}$ \cite{Ehni:2012} at this magnetic field due to chemical shielding effects from varying molecular environments. This chemical shift range is linearly spaced with $\omega_k\in[-\pi\Omega,+\pi\Omega]$ covering a bandwidth of $\Omega=40~\si{\kilo\hertz}$, slightly more than the required $37.5~\si{\kilo\hertz}$, and represents a $^{13}$C spin ensemble with $K=451$ members.

Pulses produced with the method outlined are morphed from one grid point to another in Fig.~\ref{FIG_infidelity_duration_multiplier_1}, and this morphing is key to obtaining high fidelities. This indicates that pulses should be related to each other and asks the question of whether pulses close on the grid look like each other. Twelve pulses produced by the expanded stage of the morphic optimal control method are shown in Fig.~\ref{FIG_sordor_pulses}, six of each have $\beta=\sfrac{\pi}{2}$ in Fig.~\ref{FIG_sordor_pulses}(a) and $\beta={\pi}$ in Fig.~\ref{FIG_sordor_pulses}(c), at $(200,250,300,350,400,450)\si{\micro\second}$.

Interestingly, \textsc{sordor} $\sfrac{\pi}{2}$ and \textsc{sordor} ${\pi}$ both are of very similar constitution. As shown in Fig.~\ref{FIG_sordor_pulses} a) and (c), pulse shapes in all cases are symmetric in time, while no symmetry constraints are imposed during the optimisation procedure. In essence, a parabolic time course of the pulse phase is observed, as is reminiscent of linearly frequency swept chirped pulses \cite{Baum:1985,Bohlen:1990,Bohlen:1993,Kupce:1995,Garwood:2001}, but with regular spike-like features that also frequently come with approximate $\pi$, $2\pi$ ,or $4\pi$ jumps in phase. The features are even better visualised if the quadratic phase contribution is subtracted [Figs.~\ref{FIG_sordor_pulses}(b) and \ref{FIG_sordor_pulses}(d)]. Similar phase behaviour has previously been reported for \textsc{bip} \cite{Smith:2001} and \textsc{bibop} \cite{Kobzar:2004,Kobzar:2008} $\beta=\pi$ inversion pulse shapes, but not for $\beta=\sfrac{\pi}{2}$ rotation or excitation pulses.

The comparison of the \textsc{sordor} pulses to a chirped pulse show an additional phase modulation, where the chirped phase is from the formula
\begin{align}
& \varphi(t)=\pi\Omega_\text{c}T\bigg(\frac{t}{T}-\frac{1}{2}\bigg)^2, & \Omega_\text{c}=AT & 
\end{align}
The bandwidth of this chirped phase, $\Omega_\text{c}$, is chosen to give the same pulse amplitude as the \textsc{sordor} pulses, $A=(2\pi)10~\si{\kilo\hertz}$. Clearly, this does not match the quadratic component of all \textsc{sordor} pulses in Figs.~\ref{FIG_sordor_pulses}(b) and \ref{FIG_sordor_pulses}(c), particularly for the shorter duration pulses. However, this is a good match for the $450~\si{\micro\second}$ pulses, and the comparison chosen in this way has no difference in pulse amplitude, only the extra phase modulation on top of the chirped pulse.

\textsc{Sordor} pulses are very close to symmetric, which should not be a surprise because a rotation should be a reversible operation and a time-symmetric pulse is a common form for this \cite{Ngo:1987,Kobzar:2012}. Although not implemented in this work, including this symmetry as a constraint should halve the computation time because only half the time propagators and their directional derivatives need to be calculated with Eq.~(\ref{EQ_Auxmat}).

Two \textsc{sordor} pulses are further investigated in combination: a $\sfrac{\pi}{2}$ pulse from Fig.~\ref{FIG_infidelity_duration_multiplier_1}(a), and a $\pi$ pulse from Fig.~\ref{FIG_infidelity_duration_multiplier_1}(b), each with $Q=0.78$ and $b=18$. This corresponds to a pulse duration of $T=450~\si{\micro\second}$, a time increment of $\Delta t=0.5~\si{\micro\second}$ in Eq.~(\ref{EQ_EffectivePropagators}), and the constant pulse amplitude of $A=(2\pi)10~\si{\kilo\hertz}$ in Eq.~(\ref{EQ_BilinearHamiltonian}). Both of the \textsc{sordor} pulses were obtained from the expanding stage of morphic optimal control method outlined above, these being the best performing combination of pulses produced in Fig.~\ref{FIG_infidelity_duration_multiplier_1}.

\begin{figure*}
\centering{\includegraphics{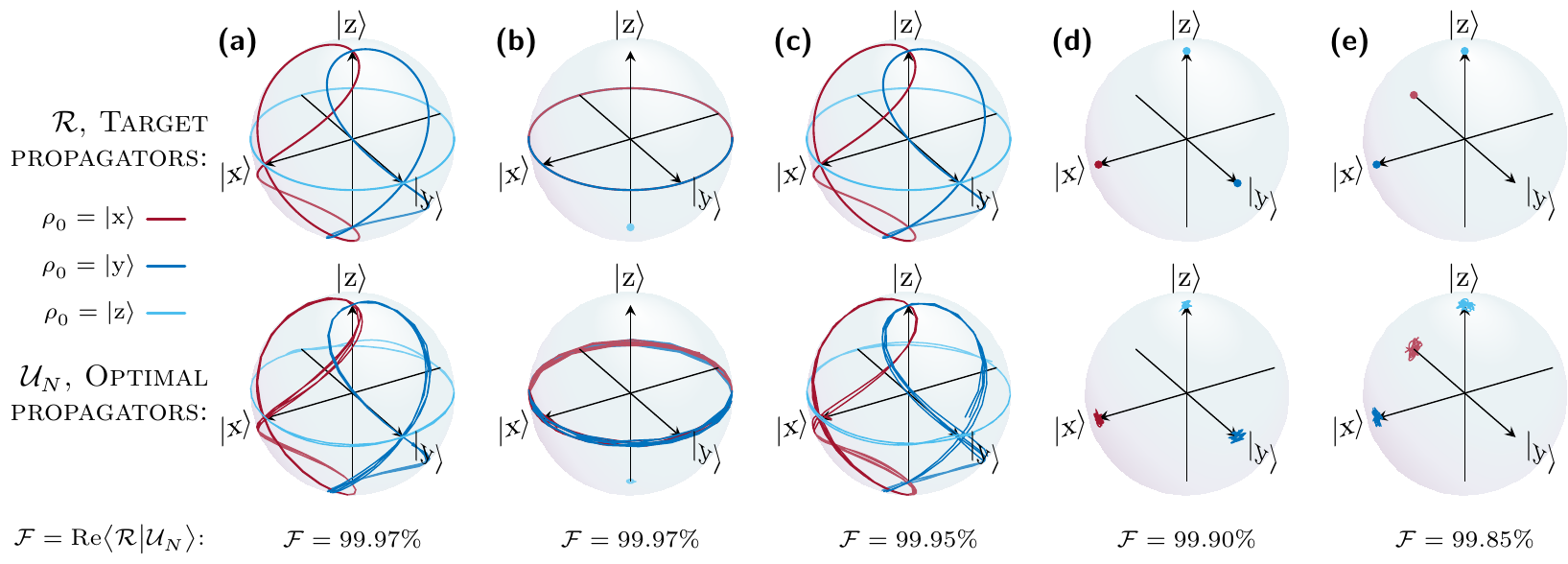}}
\caption{Plots of final states, projected onto a Bloch sphere, as a function of resonance offsets $\omega_k\in[-\pi\Omega,+\pi\Omega]$. The three line plots on each representing time propagation from initial states $\rho_0^{}=|\text{x}\rangle$ (red), $\rho_0^{}=|\text{y}\rangle$ (blue), and $\rho_0^{}=|\text{z}\rangle$ (cyan). \emph{Upper plots} show the desired unitary propagators $\mathcal{R}$. \emph{Lower plots} show the final states after propagating \textsc{sordor} pulses $\mathcal{U}_N^{}$. The columns show (a) one $\sfrac{\pi}{2}$ pulse, (b) one $\pi$ pulse, (c) the two-pulse sequence $\sfrac{\pi}{2}\to \pi$ \cite{Hahn:1950}, (d) the three-pulse sequence $\sfrac{\pi}{2}\to \pi\to \sfrac{\pi}{2}$, and (e) the five-pulse \emph{perfect echo} sequence \cite{Takegoshi:1989}.}
\label{FIG_urx_projectors_bloch}
\end{figure*}

Illustrative examples of sequences using these two \textsc{sordor} pulse are shown in Fig.~\ref{FIG_urx_projectors_bloch}. Figures \ref{FIG_urx_projectors_bloch}(a) and \ref{FIG_urx_projectors_bloch}(b) depict the performance of \textsc{sordor} $\sfrac{\pi}{2}|_\text{x}^{}$ and ${\pi}|_\text{x}^{}$ pulses (subscripts denote the axis about which an $\omega=0$ rotation is defined) projected onto a Bloch sphere for three different initial states. Figure \ref{FIG_urx_projectors_bloch}(c) depicts these two pulses in the sequence $\sfrac{\pi}{2}|_\text{x}^{}\to{\pi}|_\text{x}^{}$, named the Hahn echo \cite{Hahn:1950}, Fig.~\ref{FIG_urx_projectors_bloch}(d) is the sequence $\sfrac{\pi}{2}|_\text{x}^{}\to{\pi}|_\text{x}^{}\to\sfrac{\pi}{2}|_\text{x}^{}$, a building block within the \textsc{inept} sequence \cite{Morris:1979}, and Fig.~\ref{FIG_urx_projectors_bloch}(e) is the \emph{perfect echo} sequence $\sfrac{\pi}{2}|_\text{x}^{}\to{\pi}|_\text{y}^{}\to\sfrac{\pi}{2}|_\text{y}^{}\to{\pi}|_\text{y}^{}\to\sfrac{\pi}{2}|_{-\text{x}}^{}$ \cite{Takegoshi:1989}. The fidelities presented for each of these last three examples clearly show that \textsc{sordor} pulses can perform effectively in sequence.

\section{Conclusions}

Broadband universal rotation pulses with second order phase dispersion define a new class of pulses, named \textsc{sordor} pulses by the authors, with the universality of rotation pulses and an evolution character of adiabatic pulses. The difficulty in producing universal rotation pulses with second order phase dispersion, similar to adiabatic passage pulses, poses an interesting question regarding methods of optimal control.

A simple but effective \emph{morphic}-\textsc{grape} optimal control method has been designed to avoid local extrema traps and find shorter duration control pulses when compared with the current best universal rotation (\textsc{burbop}) pulses. The trend of this comparison shows that \textsc{sordor} pulses are approximately $50\%$ of \textsc{burbop} durations: a significant improvement.

The advance in pulse performance, compared with \textsc{burbop}, derives from a phase dispersion function used to define target unitary propagators. The particular phase dispersion function used in this work consists of a constant term, a quadratic term, and a variable multiplier. The exact form of this function is not derived within this work and is used in this form simply because the function produces high fidelity pulses. Studies using other phase dispersion functions, \emph{e.g.} of higher or arbitrary order, are left to future investigations.

\begin{acknowledgments}
The authors are grateful to Andrin Doll, Steffen Glaser, Maria Grazia Concilio, Jens Haller, Malcolm Levitt, and Herbert Ullrich for their for insightful comments and discussion. Thanks are given to Mohammadali Foroozandeh, Martin Goodwin, Ilya Kuprov, and Jason Ralph for help finalising this paper. B.L. acknowledges funding by the Deutsche Forschungsgemeinschaft (DFG LU 835/ 13-1), the HGF programme BIFTM (47.02.04), and the EU Horizon 2020 programme (667192 HYPERDIAMOND).
\end{acknowledgments}

\end{document}